\documentclass[twocolumn,twoside,slac_two]{revtex4}
\usepackage{graphicx}
\usepackage{color}
\usepackage{fancyhdr}
\input{histoc.sty}
\def\chbox#1{\hbox to 0pt{\hss #1\hss}}
\newcommand{\BF}{{\cal B}}
\newcommand{\Jpsi}{{J\mskip -3mu/\mskip -2mu\psi\mskip 2mu}}
\newcommand{\Dbar}{{\kern 0.2em\overline{\kern -0.2em D}}{}}
\newcommand{\MeV}{{\mskip 1mu\hbox{Me\kern -0.08em V}}}

\begin{document}

\title{Heavy Quark Spectroscopy}

%

\author{Roland Waldi}
\affiliation{Universit\"at Rostock, D-18051 Rostock, Germany}

\begin{abstract}
With the discovery of new states in recent years, interest in
spectroscopy has revived.  Recent experimental results in heavy
flavour spectroscopy are reviewed, including charmonium,
bottomonium, charmed mesons and baryons and bottom mesons.
\end{abstract}

\maketitle

\thispagestyle{fancy}


\section{INTRODUCTION}

Classical mesons are bound states of a quark and an anti-quark.
However, very similar properties can be expected
from states with more constituents, like glueballs, hybrids
with quark, anti-quark and (valence) gluon or
states with two quarks and two anti-quarks.

In mesons with open or hidden heavy flavour, one $q\bar q$ pair must
be present, therefore exotic mesons can only be hybrids $q\bar qg$,
meson molecules $q\bar q\,q\bar q$,
diquark bound states $qq\, \bar q\bar q$ or states with even more
constituents.  

Within the recent past, new mesons
with heavy quarks have been
observed, and several known ones have been confirmed or measured with
improved precision.
Especially
in the charmonium sector the number of new mesons found
exceeds the number of available states
in the $c\bar c$ spectrum.

The quantum numbers of new mesons can be measured in
some reactions.  Production in $e^+e^-$ annihilation either
directly or after one electron has radiated off a photon (initial state
radiation) have the quantum numbers of a photon, $J^{PC} = 1^{--}$.
Photon photon fusion leads to final states with
even $C$-parity
and in general $J^{++}\ (J \ne 1), {\it even}^{-+}$.
The quantum numbers of
states produced in two-body $B$ meson decays can be obtained
in a partial wave analysis of production and decay angles.

\section{BOTTOMONIUM SPECTROSCOPY}

\begin{figure}
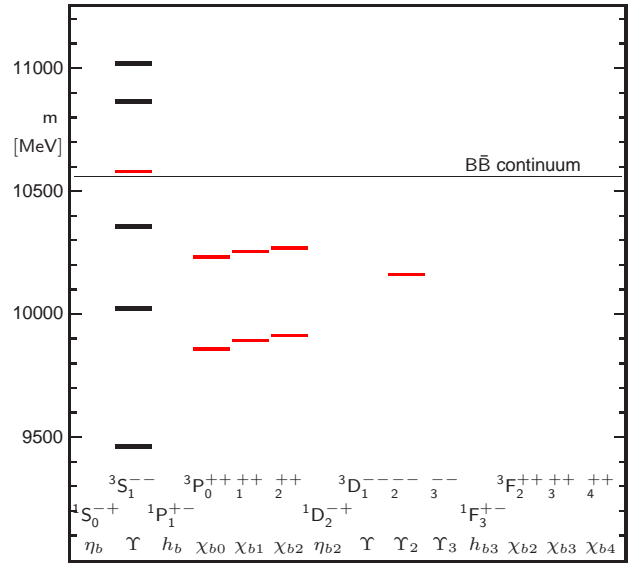

\vspace{1mm}
\hbox to \columnwidth{\hfill\sf\scriptsize
 \histogram{0.9\columnwidth}{0.9\columnwidth}{ -100}{14100}{    0}{22500}
 \histoput( 500,1600){\hbox to 0pt{\hss$\sf \ ^1S_0^{-+}$\hss}}
 \histoput(1500,1600){\raise 3ex\hbox to 0pt{\hss$\sf ^3S_1^{--}$\hss}}
 \histoput(2500,1600){\hbox to 0pt{\hss$\sf ^1P_1^{+-}$\hss}}
 \histoput(3500,1600){\raise 3ex\hbox to 0pt{\hss$\sf ^3P_0^{++}\ $\hss}}
 \histoput(4500,1600){\raise 3ex\hbox to 0pt{\hss$\sf {}_1^{++}$\hss}}
 \histoput(5500,1600){\raise 3ex\hbox to 0pt{\hss$\sf {}_2^{++}$\hss}}
 \histoput(6500,1600){\hbox to 0pt{\hss$\sf ^1D_2^{-+}$\hss}}
 \histoput(7500,1600){\raise 3ex\hbox to 0pt{\hss$\sf ^3D_1^{--}\ $\hss}}
 \histoput(8500,1600){\raise 3ex\hbox to 0pt{\hss$\sf {}_2^{--}$\hss}}
 \histoput(9500,1600){\raise 3ex\hbox to 0pt{\hss$\sf {}_3^{--}$\hss}}
 \histoput(10500,1600){\hbox to 0pt{\hss$\sf ^1F_3^{+-}$\hss}}
 \histoput(11500,1600){\raise 3ex\hbox to 0pt{\hss$\sf ^3F_2^{++}\ $\hss}}
 \histoput(12500,1600){\raise 3ex\hbox to 0pt{\hss$\sf {}_3^{++}$\hss}}
 \histoput(13500,1600){\raise 3ex\hbox to 0pt{\hss$\sf {}_4^{++}$\hss}}
 \histoput( 500,1600){\lower 3ex\hbox to 0pt{\hss$\eta_b$\hss}}
 \histoput(1500,1600){\lower 3ex\hbox to 0pt{\hss$\Upsilon$\hss}}
 \histoput(2500,1600){\lower 3ex\hbox to 0pt{\hss$h_b$\hss}}
 \histoput(3500,1600){\lower 3ex\hbox to 0pt{\hss$\chi_{b0}$\hss}}
 \histoput(4500,1600){\lower 3ex\hbox to 0pt{\hss$\chi_{b1}$\hss}}
 \histoput(5500,1600){\lower 3ex\hbox to 0pt{\hss$\chi_{b2}$\hss}}
 \histoput(6500,1600){\lower 3ex\hbox to 0pt{\hss$\eta_{b2}$\hss}}
 \histoput(7500,1600){\lower 3ex\hbox to 0pt{\hss$\Upsilon$\hss}}
 \histoput(8500,1600){\lower 3ex\hbox to 0pt{\hss$\Upsilon_2$\hss}}
 \histoput(9500,1600){\lower 3ex\hbox to 0pt{\hss$\Upsilon_3$\hss}}
 \histoput(10500,1600){\lower 3ex\hbox to 0pt{\hss$h_{b3}$\hss}}
 \histoput(11500,1600){\lower 3ex\hbox to 0pt{\hss$\chi_{b2}$\hss}}
 \histoput(12500,1600){\lower 3ex\hbox to 0pt{\hss$\chi_{b3}$\hss}}
 \histoput(13500,1600){\lower 3ex\hbox to 0pt{\hss$\chi_{b4}$\hss}}
 \histoytick{  1000}{ .7mm}{}
 \histoytick{  2000}{ .7mm}{}
 \histoytick{  3000}{ .7mm}{}
 \histoytick{  4000}{ .7mm}{}
 \histoytick{  5000}{1.4mm}{9500}
 \histoytick{  6000}{ .7mm}{}
 \histoytick{  7000}{ .7mm}{}
 \histoytick{  8000}{ .7mm}{}
 \histoytick{  9000}{ .7mm}{}
 \histoytick{ 10000}{1.4mm}{10000}
 \histoytick{ 11000}{ .7mm}{}
 \histoytick{ 12000}{ .7mm}{}
 \histoytick{ 13000}{ .7mm}{}
 \histoytick{ 14000}{ .7mm}{}
 \histoytick{ 15000}{1.4mm}{10500}
 \histoytick{ 16000}{ .7mm}{}
 \histoytick{ 17000}{ .7mm}{}
 \histoytick{ 18000}{ .7mm}{$\sf m$\ }
 \histoytick{ 18000}{  0mm}{\lower 3ex\hbox{$\sf [MeV]$}}
 \histoytick{ 19000}{ .7mm}{}
 \histoytick{ 20000}{1.4mm}{11000}
 \histoytick{ 21000}{ .7mm}{}
 \histoytick{ 22000}{ .7mm}{}
 \histobarwidth 1.5pt
 \histohline(1500+450-450, 4603)
 \histohline(1500+450-450,10233)
 \histohline(1500+450-450,13552)
{\color{red}%
 \histobarwidth 0.8pt
 \histohline(1500+450-450,15793)
 \histohline(8500+450-450,11611)
 \histohline(5500+450-450, 9122)
 \histohline(4500+450-450, 8928)
 \histohline(3500+450-450, 8594)
 \histohline(3500+450-450,12325)
 \histohline(4500+450-450,12555)
 \histohline(5500+450-450,12686)
}%
 \histohline(1500+450-450,18650)
 \histohline(1500+450-450,20190)
{\histobarwidth 0.2pt
 \histohline(0+14000-0,15600)}%
 \histoput(10000,15800){$\sf B\bar B$ \sf continuum}
\endhistogram}
\caption{The spectrum of known bottomonium states.
Data for those in red have been updated recently.}
\label{botspec}
\end{figure}

The known bottomonium spectrum is shown in Fig.~\ref{botspec}.

The vector bottomonium states were the first hadrons with $b$ quarks
discovered \cite{led77}.  Their widths were always subject
to large systematic
uncertainties.  Recent analyses by the CLEO collaboration
\cite{cleups} of the $\Upsilon(1S)$,
$\Upsilon(2S)$ and
$\Upsilon(3S)$
have improved our knowledge substantially on the essential
properties,
$\Gamma(\Upsilon \to e^+e^-)$ and
$\BF(\Upsilon \to \mu^+\mu^-)$, leading to
new and more precise values
of their total widths $\Gamma_{\rm tot}$.
A measurement by BABAR \cite{babups} has
taken into account the subtleties of a $4S$
wave function and
lead to new values of
$\Gamma(\Upsilon \to e^+e^-)$ and
$\Gamma_{\rm tot}$
of the $\Upsilon(4S)$.

Hadronic transitions
via emission of a pion pair have now also
been observed
from the $\Upsilon(4S)$ to the $\Upsilon(1S)$
and $\Upsilon(2S)$
\cite{bbpipi}.
The new transitions---added to other
hadronic transitions obeserved 
previously---reveal a peculiar pattern:
While most transitions have a $\pi\pi$ mass
spectrum peaking at the high end,
those with $\Delta n = 2$ as $3S\to1S$ and
$4S\to 2S$ show a double peak at low and
high masses.
No explanation for this effect is known to me.

The $\Upsilon(1D)$ state has already been observed
by CLEO two years ago \cite{upsd}, while other
states with $L > 1$ are still missing.

\section{CHARMONIUM SPECTROSCOPY}

The charmonium spectrum below open charm threshold
has been completed
through the confirmation of the $h_c$ state by
two experiments \cite{hc}.

\subsection{The $X,Y,Z$ States}

Many new observations in the charmonium sector have been made recently.
The oldest is $X(3872)$ observed in
$X(3872) \to \pi^+ \pi^- \Jpsi$ \cite{x3872,x3872y4260}, a narrow state
($\Gamma < 2.3\MeV$) with mass $(3871.2\pm0.4)\MeV$.  It is
produced in hadronic $B$ meson decays.
The decay rate
$X(3872) \to \gamma\Jpsi$ is lower by a factor $0.19\pm0.07$ \cite{x3872g},
and establishes even C-parity.
It has not been seen in $\gamma\gamma$ fusion and $e^+e^-$ annihilation
\cite{x3872p}.
The $\pi\pi$ mass spectrum 
peaks at the high end, consistent with
$\rho^0$ dominance. This
and a full angular analysis of
$X(3872) \to \pi^+ \pi^- \Jpsi$ \cite{x3872a}
rule out all assignments except $J^{PC} = 1^{++}$.
That would be $\chi_{c1}$, but there is no vacant slot
in the $c\bar c$ spectrum for such a state.

As an exotic state, its isospin might be
different from 0.  In fact, the transition
$\chi_{c1}(nS) \to \pi^+ \pi^- \Jpsi$ is violating isospin conservation,
and can therefore be no strong transition.
However, no charged partner
is observed in $B$ decays \cite{x3872b}.
On the other hand, there is some indication
of the isospin-allowed
$X(3872) \to \omega\Jpsi$ \cite{x3872o}, although kinematics
allows only $\pi^+\pi^-\pi^0$ below the 
central value of the $\omega$
mass.

An inclusive search in $B$ decays to determine the absolute branching
fractions of the $X(3872)$ was not (yet) successful \cite{x3872i}.

A peak at threshold in $D^0\Dbar^0\pi^0$ at $3875\MeV$
\cite{x3872d}
is probably the same $X(3872)$ state.

Since it does not fit into the $c\bar c$ spectrum, the $X(3872)$ is
an exotic meson with $J^{PC} = 1^{++}$ and most likely $I=0$.

Another new state has been observed in $e^+ e^- \to c\bar c c\bar c$-production:
$e^+e^- \to \Jpsi\,X(3940)$, $X(3940) \to D^* D$ \cite{x3940}.
It is probably narrow with a measured $\Gamma = (39\pm26)\MeV$,
and is not seen in the $\Dbar D$ final state.

A second state at the same mass has been observed in $B$ decays,
which seems to be broader with
$\Gamma = (87\pm33)\MeV$ and decays into $\omega \Jpsi$ \cite{y3940}.
It is not seen in either $\Dbar D$ or $\Dbar^* D$.  All these
properties distinguish it from the $X(3940)$,
so it has been called $Y(3940)$.

A third state at the same or a little bit smaller
mass has been called $Z(3930)$.  It is produced in
$\gamma\gamma$ fusion and decays to $\Dbar D$ \cite{z3930}.
The helicity angle distribution favors spin 2, so it is likely
a $2^{++}$ meson.  Its mass would fit the expected
$c\bar c$-meson $\chi_{c2}(2P)$ so it is the least peculiar of the new
mesons.

At a somewhat higher mass, another state was observed in radiative
$e^+e^-$ annihilation
$e^+e^- \to \gamma Y(4260)$ decaying into
$\pi^+ \pi^i \Jpsi$ \cite{y4260}.
A scan by CLEO-c \cite{y4260c} reveals an enhancement
at direct production from $e^+e^-$ annihilation at this energy,
with a ratio
$\pi^+ \pi^- \Jpsi$ to
$\pi^0 \pi^0 \Jpsi$ of approximately $2:1$ as expected for a
$I=0$ hadronic transition between vector charmonium states.
However,
vector charmonium states are expected to show up in
the inclusive hadronic cross section of $e^+e^-$ annihilation
(in Fig.~40.7 in the PDG book \cite{pdg06}),
but at $\sqrt s = 4260\MeV$ there is a dip!

The state has been possibly observed
at the $3\sigma$ level in hadronic $B$
decays \cite{x3872y4260}.
The decay to $\Dbar D$ ist not observed and is less than $7.6\%$ of
$\pi^+ \pi^- \Jpsi$ at 95\% CL \cite{y4260d}.

There is also a state at mass $4320\MeV$ decaying into
$\pi^+ \pi^- \psi(2S)$ \cite{y4320}.

In summary, several new states have been observed recently that do
not match with available slots in the $c\bar c$ spectrum.
At least 4 of these states have to be considered exotic mesons.

As of their true nature, there exist many speculations \cite{xyztheory},
but no conclusive arguments have been found for a unique solution for
any of these.

\section{OPEN FLAVOUR MESONS}

In contrast to quarkonia,
the spectrum of heavy mesons with open flavour is
given by eigenstates to $L$ and $j_q$, the total angular momentum
(orbital $L$ plus spin $S_q$) of the light quark,
rather than $L$ and $S$, the sum $S_q$ plus $S_Q$ of
both quark spins.  The corresponding notation $^{(j)}L^P_J$
is used in figure~\ref{dsspec}.

\subsection{$B$, $B_s$ and $B_c$ Mesons}

There are recent measurements of excited $B$ mesons
$B_J^0 \to B^{(*)+} \pi^-$ \cite{bj,talkteva}.  Unfortunately, the masses
reported by CDF and D0 differ significantly, and more data are
needed to get at precise masses and widths.

D0 has also observed evidence for the
$B_{s2}^*$ meson
at a mass  of $(5839.1\pm1.4\pm1.5)\MeV$
\cite{bsj}.
This value has been confirmed after this conference
by CDF with higher precision \cite{bsjcdf}.

It is also worthwhile to mention that a clear peak
of 39 fully reconstructed events of the double-heavy
$B_c$ meson has been presented this year \cite{bc}.

\subsection{$D_s$ Mesons}

\begin{figure}
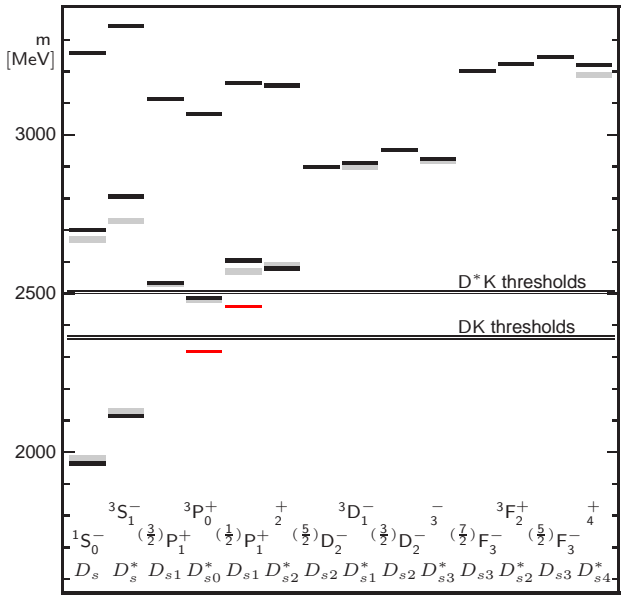

\vspace{1mm}
\hbox to \columnwidth{\hfill\sf\scriptsize
 \histogram{0.9\columnwidth}{0.95\columnwidth}{ -100}{14100}{15600}{34000}
 \histoput( 500,17000){\hbox to 0pt{\hss$\sf \ ^1S_0^{-}$\hss}}
 \histoput(1500,17000){\raise 3ex\hbox to 0pt{\hss$\sf ^3S_1^{-}$\hss}}
 \histoput(2500,17000){\hbox to 0pt{\hss$\sf ^{({3\over2})}P_1^{+}$\hss}}
 \histoput(3500,17000){\raise 3ex\hbox to 0pt{\hss$\sf ^3P_0^{+}\ $\hss}}
 \histoput(4500,17000){\hbox to 0pt{\hss$\sf ^{({1\over2})}P_1^{+}$\hss}}
 \histoput(5500,17000){\raise 3ex\hbox to 0pt{\hss$\sf {}_2^{+}$\hss}}
 \histoput(6500,17000){\hbox to 0pt{\hss$\sf ^{({5\over2})}D_2^{-}$\hss}}
 \histoput(7500,17000){\raise 3ex\hbox to 0pt{\hss$\sf ^3D_1^{-}\ $\hss}}
 \histoput(8500,17000){\hbox to 0pt{\hss$\sf ^{({3\over2})}D_2^{-}$\hss}}
 \histoput(9500,17000){\raise 3ex\hbox to 0pt{\hss$\sf {}_3^{-}$\hss}}
 \histoput(10500,17000){\hbox to 0pt{\hss$\sf ^{({7\over2})}F_3^{-}$\hss}}
 \histoput(11500,17000){\raise 3ex\hbox to 0pt{\hss$\sf ^3F_2^{+}\ $\hss}}
 \histoput(12500,17000){\hbox to 0pt{\hss$\sf ^{({5\over2})}F_3^{-}$\hss}}
 \histoput(13500,17000){\raise 3ex\hbox to 0pt{\hss$\sf {}_4^{+}$\hss}}
 \histoput( 500,17000){\lower 3ex\hbox to 0pt{\hss$D_s$\hss}}
 \histoput(1500,17000){\lower 3ex\hbox to 0pt{\hss$D_s^*$\hss}}
 \histoput(2500,17000){\lower 3ex\hbox to 0pt{\hss$D_{s1}$\hss}}
 \histoput(3500,17000){\lower 3ex\hbox to 0pt{\hss$D^*_{s0}$\hss}}
 \histoput(4500,17000){\lower 3ex\hbox to 0pt{\hss$D_{s1}$\hss}}
 \histoput(5500,17000){\lower 3ex\hbox to 0pt{\hss$D^*_{s2}$\hss}}
 \histoput(6500,17000){\lower 3ex\hbox to 0pt{\hss$D_{s2}$\hss}}
 \histoput(7500,17000){\lower 3ex\hbox to 0pt{\hss$D^*_{s1}$\hss}}
 \histoput(8500,17000){\lower 3ex\hbox to 0pt{\hss$D_{s2}$\hss}}
 \histoput(9500,17000){\lower 3ex\hbox to 0pt{\hss$D^*_{s3}$\hss}}
 \histoput(10500,17000){\lower 3ex\hbox to 0pt{\hss$D_{s3}$\hss}}
 \histoput(11500,17000){\lower 3ex\hbox to 0pt{\hss$D^*_{s2}$\hss}}
 \histoput(12500,17000){\lower 3ex\hbox to 0pt{\hss$D_{s3}$\hss}}
 \histoput(13500,17000){\lower 3ex\hbox to 0pt{\hss$D^*_{s4}$\hss}}
 \histoytick{ 16000}{ .7mm}{}
 \histoytick{ 17000}{ .7mm}{}
 \histoytick{ 18000}{ .7mm}{}
 \histoytick{ 19000}{ .7mm}{}
 \histoytick{ 20000}{1.4mm}{2000}
 \histoytick{ 21000}{ .7mm}{}
 \histoytick{ 22000}{ .7mm}{}
 \histoytick{ 23000}{ .7mm}{}
 \histoytick{ 24000}{ .7mm}{}
 \histoytick{ 25000}{1.4mm}{2500}
 \histoytick{ 26000}{ .7mm}{}
 \histoytick{ 27000}{ .7mm}{}
 \histoytick{ 28000}{ .7mm}{}
 \histoytick{ 29000}{ .7mm}{}
 \histoytick{ 30000}{1.4mm}{3000}
 \histoytick{ 31000}{ .7mm}{}
 \histoytick{ 32000}{ .7mm}{}
 \histoytick{ 33000}{ .7mm}{$\sf m$\ }
 \histoytick{ 33000}{  0mm}{\lower 2ex\hbox{$\sf [MeV]$}}
\definecolor{gray}{rgb}{0.8,0.8,0.8}%
{\color{gray}%
 \histobarwidth 2pt
 \histohline(500+450-450,19800)
 \histohline(500+450-450,26700)
 \histohline(1500+450-450,21300)
 \histohline(1500+450-450,27300)
 \histohline(2500+450-450,25300)
 \histohline(3500+450-450,24800)
 \histohline(4500+450-450,25700)
 \histohline(5500+450-450,25900)
 \histohline(7500+450-450,29000)
 \histohline(9500+450-450,29200)
 \histohline(13500+450-450,31900)
}%
{%
 \histobarwidth 1.2pt
 \histohline(500+450-450,19650)
 \histohline(500+450-450,27000)
 \histohline(500+450-450,32590)
 \histohline(1500+450-450,21130)
 \histohline(1500+450-450,28060)
 \histohline(1500+450-450,33450)
 \histohline(4500+450-450,26050)
 \histohline(4500+450-450,31650)
 \histohline(3500+450-450,24870)
 \histohline(3500+450-450,30670)
 \histohline(2500+450-450,25350)
 \histohline(2500+450-450,31140)
 \histohline(5500+450-450,25810)
 \histohline(5500+450-450,31570)
 \histohline(8500+450-450,29530)
 \histohline(7500+450-450,29130)
 \histohline(6500+450-450,29000)
 \histohline(9500+450-450,29250)
 \histohline(12500+450-450,32470)
 \histohline(11500+450-450,32240)
 \histohline(10500+450-450,32030)
 \histohline(13500+450-450,32200)
}%
{\color{red}%
 \histobarwidth 0.8pt
 \histohline(3500+450-450,23173)
%
 \histohline(4500+450-450,24589)
}%
 \histohline(2500+450-450,25354)
 \histohline(5500+450-450,25735)
{\histobarwidth 0.2pt
 \histohline(0+14000-0,23583)
 \histohline(0+14000-0,23670)
 \histoput(10000,23770){$\sf DK$ \sf thresholds}
 \histohline(0+14000-0,25004)
 \histohline(0+14000-0,25076)
 \histoput(10000,25176){$\sf D^* K$ \sf thresholds}
}%
\endhistogram}
\caption{The spectrum of $c\bar s$ states.
Theoretical predictions are from
\cite{gi} (grey) and
\cite{dipei} (black).
The puzzling new states are the thin lines in red.}
\label{dsspec}
\end{figure}

A strong decay of excited $D_s$ mesons
has to conserve isospin and thus goes to
$D^{(*)} K$ final states.
The spectrum in figure~\ref{dsspec}
shows the expected states and the
$D^{(*)} K$ thresholds.

Recently observed new states
$D_{s0}^*(2317)$ and
$D_{s1}^*(2460)$ \cite{dsj}
are both below the threshold of their allowed
decay mode, and have been observed decaying into
$D_s^{+} \pi^0$ and
$D_s^{*+} \pi^0$, respectively.
Consistently, these states are narrow with
upper limits of a few $\MeV$ for their widths.

Their spin-parity assignments are consistent with their
observation (or non-observation) in the transitions to
$D_s \pi^0$, 
$D_s \gamma$, 
$D_s^* \pi^0$, 
and $D_s \pi^+\pi^-$ \cite{dsjt},
and with their decay angular distributions \cite{dsjd}. 
What is puzzling about these mesons is that
all
model calculations (unless heavily tweaked)
predict significantly larger
masses for these states in the $c\bar s$ system which makes them
also good candidates for exotic mesons.

Another state
$D_{sJ}(2860)$ has been observed recently by BABAR \cite{dsj2860}
which decays into $D K$.  Although it has been suggested to be a 
radial excitation
of the exotic $D_{s0}^*(2317)$ meson, it could also be
an ordinary $c\bar s$ meson.

\section{CHARMED BARYONS}

Several new excited baryons have been added to the
list of known hadrons recently.
They fill up the existing open slots in the
spectrum of heavy baryons.

In the $cud$ system, the
$\Lambda_c^+(2880)$ \cite{lc} and a new state
$\Lambda_c^+(2940)$ have been observed 
to decay into $D^0 p$ \cite{lcdp}.
Their assignment $I=0$ is supported by the absence of
a partner with the $D^+ p$ final state.

There is a triplet of $I=1$ mesons
$\Sigma_c^+(2800)$ decaying to $\Lambda_c^+ \pi$ \cite{sc2800}.

New $c u s$ baryons
$\Xi_c^+(2980)$ and
$\Xi_c^+(3077)$ decaying to $\Lambda_c^+ K^-\pi^+$ \cite{xc}
have also been seen, and there is
evidence for their isodoublett partners
$\Xi_c^0(2980)$ and
$\Xi_c^0(3077)$ decaying to $\Lambda_c^+ K^0_S \pi^+$ \cite{xc0}.

\section{SUMMARY}

The appearance of states in excess to the $c\bar c$ spectrum
indicates clearly the presence of exotic mesons which are made of
more constituents than just a quark and an anti-quark.
The interpretation of the new states is a challenge to
non-perturbative QCD calculations, which will
hopefully be able to calculate the spectrum of ordinary and exotic
mesons, and thus be able to predict the
masses and widths of by now undiscovered states.

\subsection*{Acknowledgments}

I have much enjoyed the personal atmosphere
of the conference.
In preparing this talk, I have profited from talks and material of
many colleagues at BABAR as well as from input by members of other
experiments.

\end{document}